# A Novel Thermal Network Model and Electro-Thermal Coupling Study for NSFETs and CFETs Considering Thermal Crosstalk

Tianci Miao[1], Qihang Zheng[1], Yangyang Hu[1], Xiaoyu Cheng[1], Jie Liang[1], Liang Chen[1,2,3], Aiying Guo[1,2,3,4], Jingjing Liu[1,2,3], Kailin Ren[1,2,3,4], *, and Jianhua Zhang[1,2,3,4]

1  School of Microelectronics, Shanghai University, Shanghai 200444, China

2  Shanghai Collaborative Innovation Center of Intelligent Sensing Chip Technology, Shanghai University, Shanghai 200444, China

3  Shanghai Key Laboratory of Chips and Systems for Intelligent Connected Vehicle, Shanghai University, Shanghai 200444, China

4  Key Laboratory of Advanced Display and System Applications (Ministry of Education), Shanghai University, Shanghai 200444, China

*  Correspondence: renkailin@shu.edu.cn

**Abstract—As the technology node continues to shrink, nanosheet field effect transistors (NSFETs) and complementary FETs (CFETs) become valid candidates for the 3nm and sub-nanometre nodes. However, due to the shrinking device size, self-heating and inter-device thermal crosstalk of NSFETs and CFETs become more severe. It is important to accurately calculate the self-heating and thermal crosstalk of devices and to study the electrical and thermal characteristics of logic gates, etc. In this work, a thermal network model considering the thermal crosstalk of neighboring devices is proposed, which can accurately calculate the self-heating and thermal crosstalk. The electrical and thermal characteristics of NSFETs and CFETs are compared, and it is found that CFETs have more severe self-heating and thermal crosstalk. The electro-thermal characteristics of inverters, logic gates and ring oscillators composed of NSFETs and CFETs are further investigated. Compared with NSFETs, logic gates and ring oscillators composed of CFETs are more seriously affected by self-heating and should be given extra attention. The thermal network model proposed in this paper can be further used to study the thermal optimization strategy of devices and circuits to enhance the electrical performance, achieving the design technology co-optimizations (DTCO).**

## I.  Introduction

Nanosheet field effect transistors (NSFETs) have been widely studied [2], [3], owing to their superior gate controllability and adjustable channel widths [3], which offers significant advantages from a circuit design point of view[3].

However, the spacing between NFETs and PFETs in NSFETs limits further scaling of cell height. Unlike NSFETs, complementary FETs (CFETs) employ a vertical stacking structure of NFETs and PFETs, which can avoid this problem and further increase the chip integration density by a factor of 1.5-2 [4], [5].

However, when the thickness of the material is reduced to the nanometre scale, the probability of phonon collisions increases, and the enhanced scattering effect degrades the thermal conductivity of the material and intensifies the self-heating effect (SHE) [6]. Moreover, the increasing integration density and the use of low thermal conductivity materials lead to more severe SHE. The thermal crosstalk between neighboring devices is also more severe due to the decreasing device spacing, so CFETs have more severe thermal crosstalk due to the top and bottom stacked structure. The SHE will lead to on-state current degradation, threshold voltage shift, etc. At the same time, the reliability related to hot carrier injection (HCI) and bias temperature instability (BTI) will be deteriorated [7], [8]. Therefore, it is of great significance to study the SHE of NSFETs and CFETs, to establish accurate self-heating models, and to investigate the electro-thermal coupling characteristics of devices and circuits.

At present, there are two main approaches to self-heating and electro-thermal coupling research: simulation and modelling. For simulation, Cai et al. used finite element tools to study the effect of NSFET structural parameters on self-heating [9]. Myeong et al. found that the inter-layer metal thickness has a large effect on the RC delay and self-heating of NSFETs [10]. Liu et al. proposed a new type of side-wall structure that mitigates the self-heating without sacrificing the electrical performance [11]. However, the time required to study the self-heating and electro-thermal coupling of the device by simulation is long and inefficient. In terms of modelling, Cai et al. proposed a physical based thermal network model for NSFETs, however, this model is more difficult to integrate into the BSIM-CMG model [12]. Liu et al. proposed a thermal resistance model considering thermal coupling between the channels of NSFETs [13]. Zhao et al. fitted the thermal transient response curves of CFETs, and proposed a cross-coupled thermal network model considering the thermal crosstalk between the N/PFETs, however, there is no further CFET to study the electro-thermal coupling characteristics [14]. However, there is no model that can accurately calculate the crosstalk of neighboring devices in electro-thermal coupling calculations. In this work, a thermal network model considering the thermal crosstalk of neighboring devices is proposed, which can accurately calculate the self-heating and thermal crosstalk.

This article is organized as follows. Section II describes the device structures of NSFETs and CFETs, the overall research flow of this paper and TCAD simulation results. In Section III, based on the finite element modeling (FEM) simulation results, a thermal network model considering thermal crosstalk between neighboring devices is proposed. In Section IV, the proposed model is integrated into the BSIM-CMG model to investigate the electro-thermal characteristics of inverters, logic gates and ring oscillators (RO) composed of the two devices. Finally, the paper is concluded in Section V.

## II. DEVICE STRUCTURE AND SIMULATION RESULTS

### A. Device Structure

Fig. 1 illustrates the structure of NSFET and CFET with BPR. The top device of CFET is NFET, the bottom device is PFET, M0 is a wrap-around contact (WAC) structure [15], [16], M0 is tungsten (W), the BPR and via BPR (Vbpr) are ruthenium (Ru) [17], and high-k isolation layer between the BPR and the substrate is SiO2. The structural parameters of the device are referred to in literature [18], the NSFET is designed as A14 (42 nm CPP, 18 nm MP), and the CFET is designed as A7 (39 nm CPP, 16 nm MP), the detailed structural parameters are shown in Table I.

The thermal conductivity and constant pressure heat capacity of different materials at 300 K are listed in Table II [10], [19], [20]. Based on the literature [21], [22] and [23], it can be concluded that temperature, thickness and doping concentration have a significant effect on the thermal conductivity of Silicon (Si) and SiliconGermanium (SiGe). Therefore, this effect is taken into account in calculating the thermal conductivity of Si and SiGe in this study. The thermal conductivity of Si was calculated from reference [21] and the thermal conductivity of SiGe was calculated from references [23], [24].

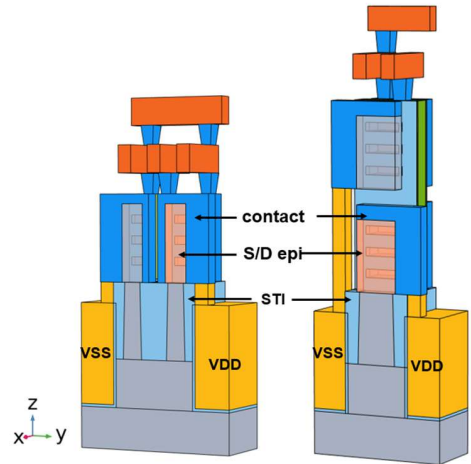

Fig. 1. Three-dimensional structures of NSFET (left) and CFET (right).

TABLE I
**STRUCTURAL PARAMETERS**

| Symbol | Parameter | Value [nm] |
|---|---|---|
| $L_g$ | Gate Length | 12 |
| $W_{ns}$ | Nanosheet Width | 12 (NSFET) 25 (CFET) |
| $H_{ns}$ | Nanosheet Thickness | 5 |
| $L_{sd}$ | Source/Drain Length | 10 |
| $H_{BPR}$ | Height of BPR | 70 |
| $H_{SiO2}$ | Gate Oxide Thickness | 0.5 |
| $H_{Hf02}$ | High-k Gate Oxide Layer | 1.5 |
| $N_{ch-N}$ | Channel doping concentration in NFET (B) | $1 \times 10^{16}$ cm$^{-3}$ |
| $N_{ch-P}$ | Channel doping concentration in PFET (As) | $1 \times 10^{16}$ cm$^{-3}$ |
| $N_{S/D-N}$ | S/D region doping concentration in NFET (As) | $5 \times 10^{20}$ cm$^{-3}$ |
| $N_{S/D-P}$ | S/D region doping concentration in PFET (B) | $5 \times 10^{20}$ cm$^{-3}$ |

TABLE II
**MATERIAL THERMAL PROPERTIES**

| Material | k (W/(K·m)) | $C_p$ (J/(Kg·K)) |
|---|---|---|
| Channel | 10 | 714 |
| Source/Drain (NFET) | 2.2 | 714 |
| Source/Drain (PFET) | 0.67 | 642 |
| SiO2 | 1.4 | 301 |
| Metal Gate (TiN) | 19.2 | 224 |
| Si bulk | 148 | 714 |
| BPR (Ru) | 115 | 238 |
| Mint (Cu) | 400 | 383 |
| Contact (W) | 175 | 132 |

The overall research flow of this paper is shown in Fig. 2. TCAD simulation is used to obtain the CV and IV curves and the power consumption of the device. The thermal characteristics of the device are simulated using FEM, the equivalent thermal network model is established, and the parameters of the BSIM-CMG model are extracted with and without self-heating respectively in conjunction with the established equivalent thermal network model, and then the electro-thermal characteristics of the inverters, logic gates, and ring oscillators are investigated.

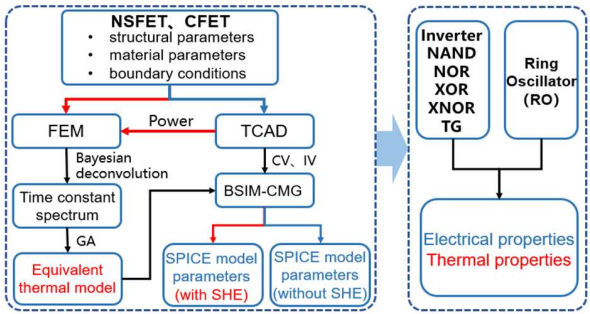

Fig.2. Thermal network modelling and electro-thermal coupling research flow in this work.

## B. Simulation Setups and Results

The electro-thermal characteristics of NSFET and CFET are investigated using the Sentaurus TCAD tool [25]. Thermodynamic model is applied when the SHE is considered. To reflect the quantum confinement effect, the density-gradient model is incorporated. The Slotboom bandgap narrowing model is considered for the doping dependent bandgap narrowing in all the device regions. The mobility models include Thin layer, Phumob, and Enormal (Lombardi) to account for Coulomb scattering, interfacial surface roughness scattering, and mobility degradation at high field. The deformation potential model with two-band k · p model for electrons and six-band k · p model for holes is used to account for the stress induced changes in bandgaps, effective masses, and effective density of states. The influence of stress on mobility is represented by the mobility enhancement factor.

Fig. 3(a) represent the calibration of the simulated and experimental data for NSFETs considering the SHE at $V_{DS} = 0.7$ V [26]. Fig. 3(b) shows the simulation results of the two devices NFET and PFET with and without self-heating. The CFET has a larger on-state current and a larger current degradation due to its larger nanosheet width. Due to the lower zero temperature coefficient point of PFET, its on-state current degradation with SHE is greater than that of NFET.

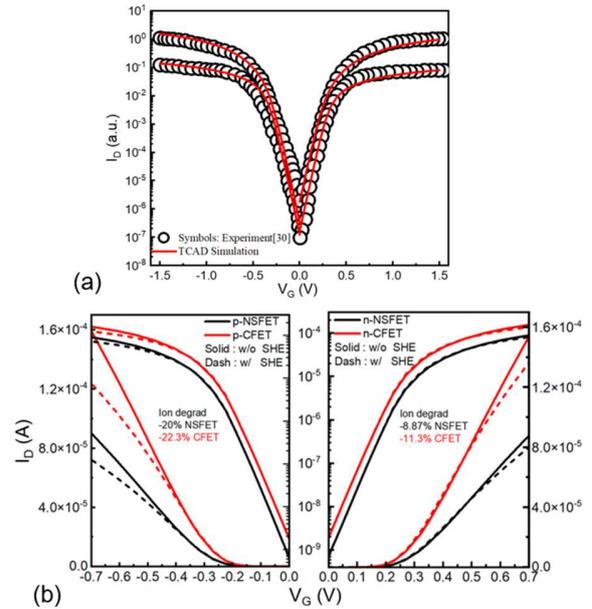

Fig. 3. (a) $I_{DS}$-$V_{GS}$ calibration results for NSFETs, (b) $I_{DS}$-$V_{GS}$ curves obtained by TCAD simulations of NSFET and CFET.

## III. MODELLING OF THERMAL NETWORKS CONSIDERING THERMAL CROSSTALK OF ADJACENT DEVICES

## A. FEM Simulation

According to the $I_d V_g$ curves in Fig. 3(b), the input heat source for the FEM is set to be $I_{on} \times V_{DD}$, and the heat sources are 56 μW and 50.5 μW for the NFET and PFET in NSFET, and 95.4 μW and 86.6 μW for the CFET, respectively.The thermal conductivities of the materials are set up according to the Table II, and the finite element simulation software COMSOL is used to simulate the thermal characteristics of the NSFETs and CFETs. There is an ideal heat sink between the substrate and back end of line (BEOL), and the rest of the boundaries are set to be thermally isolated to external objects. The simulation results in Fig. 4 illustrate the temperature distribution when the NFET is turned on. The maximum temperature is 433 K for the NSFET (a) and 534 K for the CFET (b). The highest temperatures are all in the channel near the drain. The temperature rise of the PFET is 441 K for the NSFET, and that of the CFET is 517 K. For the PFET, due to the use of SiGe with low thermal conductivity for the source/drain, it has a greater temperature rise when it has the same heat source size as the NFET.

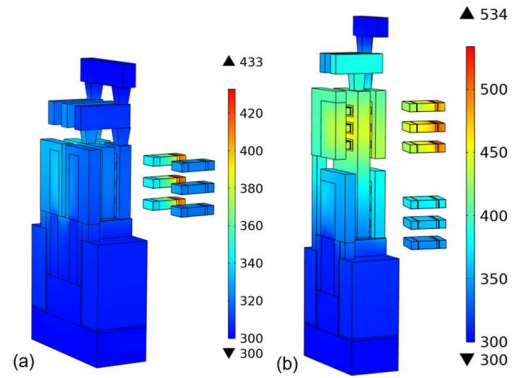

Fig. 4. Temperature distribution obtained by FEM when the NFET is on, (a) NSFET, (b) CFET.

The thermal crosstalk between neighboring devices is also more pronounced due to the reduced device pitch, and the thermal crosstalk between N/PFET is more severe for CFETs due to the top and bottom stacked structure. The device thermal crosstalk coefficients are calculated as shown in equation (1). $\Delta T_{pn}$ is the thermal crosstalk of the PFET to the NFET, $\Delta T_{jp}$ is the self-heating of the PFET, and $P_p$ is the power consumption of the PFET The thermal crosstalk coefficient is 17% for NSFET and 32% for CFET. It is shown that the thermal crosstalk between the devices is quite serious and should be taken into account when conducting electro-thermal coupling studies.

$$\rho = \frac{\Delta T_{pn}/P_p + \Delta T_{np}/P_n}{\Delta T_{jp}/P_p + \Delta T_{jn}/P_n} \qquad (1)$$

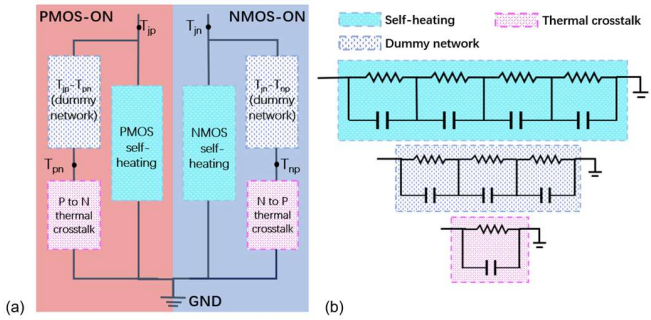

Fig.5. (a) The proposed thermal network model considering self-heating and thermal crosstalk of neighboring devices, and (b) the corresponding thermal resistance-heat capacity network for each part of the model.

### B. Model Building

The proposed thermal network model considering thermal crosstalk between neighboring devices is shown in Fig. 5, which is divided into two parts, NFET-ON and PFET-ON, and each part contains device self-heating as well as the effect on neighboring devices. In addition,（断开句子）a dummy thermal network is also added to balance the power, and Fig. 5(b) illustrates the corresponding thermal resistance and capacitance network for each part in Fig. 5 (a). For the extraction of thermal resistance and heat capacity of each part, firstly, FEM is used to obtain the thermal transient response curve, and Bayesian deconvolution is used to obtain the time-constant spectra, and the number of peaks in the time-constant spectra is used to determine the order of the thermal network, and finally the thermal transient response curve is fitted using genetic algorithm (GA) to obtain the thermal resistance heat capacity values. Fig. 6(a) shows the time constant spectrum obtained by Bayesian deconvolution of the thermal transient response curve of the NFET in the NSFET, and the results of the FEM fit to the model of the thermal transient response curve of the NFET during conduction is depicted in Fig. 6(b).

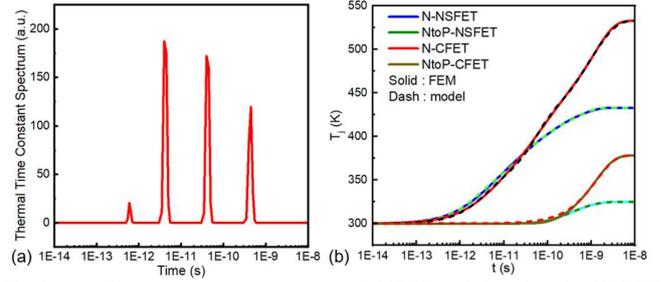

Fig.6. (a) Time constant spectrum of NFET self-heating in NSFET obtained by Bayesian back-convolution, (b) Thermal transient response curves of the model and FEM fitting when the NFET is on.

### C. SPICE Model Parameter Extraction

In order to study the electrical and thermal characteristics of inverters and logic gates, it is necessary to extract the BSIM-CMG model parameters, and the parameter extraction flow of this paper is illustrated in Fig. 7 [27]. Among them, Step 1~Step 4 are the extraction of the basic electrical parameters of the device when the SHE is not considered. On the basis of Step 1~Step 4, Step 5~Step 7 are the extraction of the parameters when the device considers the SHE, and the extraction is done by turning on the switch of the SHE in the BSIM-CMG model (SHMOD=1), as well as by integrating the proposed thermal network model into the model to accurately compute the temperature of the device, so that the model parameters can be obtained when the SHE is considered in the end. The results of fitting the NFET model with TCAD for NSFET are shown in Fig. 8, (a)-(c) for the fitting without SHE, and (d)-(f) for the fitting with SHE, and the overall root mean square error of the fitting is less than 3.5%.

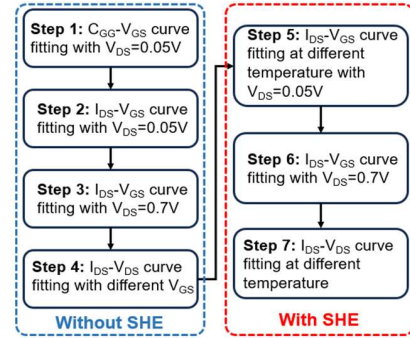

Fig.7. Extraction process of BSIM-CMG electrothermal parameters considering SHE.

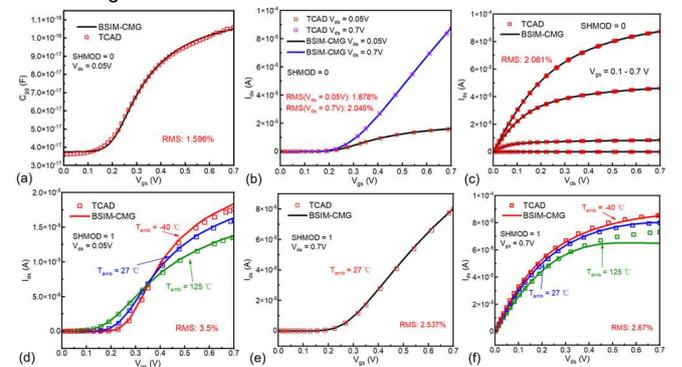

Fig.8. Fitting results of the BSIM-CMG model and TCAD simulation of the NFET in NSFET, (a) - (c) fits without SHE, (d) - (f) fits with SHE.

## IV. ELECTROTHERMAL COUPLING STUDY OF INVERTERS, LOGIC GATE, RO

### A. Simulation Of Electrical And Thermal Characteristics Of Inverters

After obtaining the parameters of the BSIM-CMG model, the electro-thermal characteristics of the inverter were firstly investigated. In the transient simulation, for the input signal Vin, both the rising and falling edge times are set to 1 ps and the period is set to 10 ns. As shown in Fig. 9, the effect of self-heating on the transmission delay of the NSFET and CFET inverters is investigated for different load capacitances ($C_L$). As shown in Fig. 10(a), with the increase of $C_L$, the charging and discharging current and time of the device increase, the SHE of the device is intensified and the temperature rises, so the impact on the electrical characteristics of the inverter is greater, as illustrated in Fig. 10 (a), when $C_L$ is 2 fF, the operating temperatures of the NSFET's NFET and PFET increase by 40 K and 40.3 K, respectively, and the impact on the transmission delay is caused by less than 1%; When $C_L$ is 20 fF, the NFET and PFET rise up to 74 K and 80.5 K, respectively, which increases the transmission delay by 6.7%.

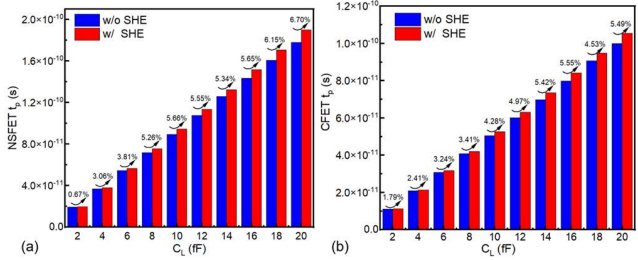

Fig.9. Inverter transmission delay with/without SHE at different load capacitances $C_L$ and effect of SHE on transmission delay for (a) NSFET and (b) CFET.

It can also be seen from Fig. 9 that the effect of self-heating on the transmission delay of the CFET is slightly smaller than that of the NSFET. The probable reason for this is that the CFET has a larger on-state current, and hence its switching on and off time are smaller compared to NSFET, resulting in a transient current slightly smaller than that of the NSFET as shown in Fig. 10(b), and hence the corresponding temperature rise is slightly smaller than that of the NSFET. As shown in Fig. 10(a), when $C_L$ is 20 fF, the temperature rise of the NFET and PFET of the CFET is 68.3 K and 75.4 K, and the temperature rise of the NSFET is 74 K and 80.5 K, respectively.

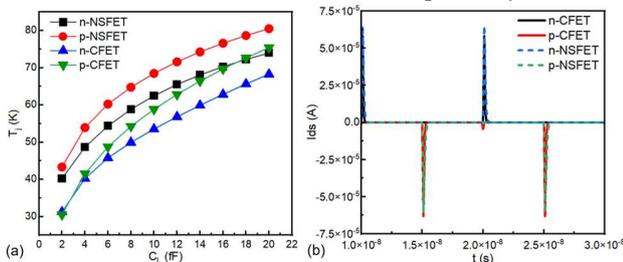

Fig.10. (a) Transient simulated temperature rise of NSFET and CFET inverters with different load capacitance $C_L$, (b) Transient currents of NSFET and CFET when $C_L$ is 20fF.

### B. Simulation Of Electrical And Thermal Characteristics Of Logic Gate

In digital logic circuits, apart from inverters, other logic units are also critical for the overall circuit. In this paper, transient simulation of basic logic units such as TG、NAND、NOR、XOR and XNOR is also carried out in order to study the effect of SHE on the rise time, and fall time of the above mentioned basic logic gates. In the simulation, both the rising and falling edge times are set to 1 ps and the period is set to 10 ns, and the $C_L$ is set to 20 fF.

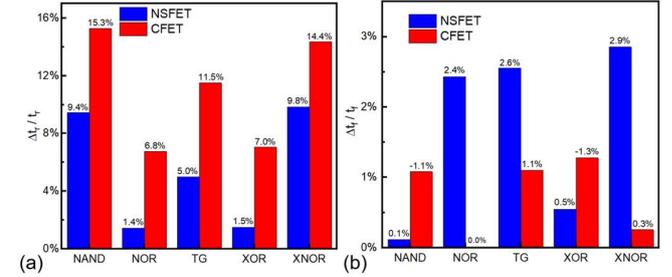

Fig.11. Variation of (a) rise time (tr) and (b) fall time (tf) caused by SHE in various logic gates consisting of NSFET and CFET with $C_L$ of 20fF.

The rise time and fall time of the logic cell are affected by the SHE is depicted in Fig. 11. For the rise time, self-heating has a greater effect on CFETs than on NSFETs, while for the fall time, it is the opposite, and because the zero- temperature coefficient point of NFETs is further back than that of PFETs, self-heating of certain logic gates of CFETs will instead shorten the fall time. NAND as an example, when the output signal is converted from 0 to 1, NFET turn off, PFET at least one turn on, the power supply voltage through the PFET device to the load capacitance charging, due to the PFET for the parallel relationship, the source-drain voltage is larger, so it is subjected to SHE is greater; when the output signal is converted from 1 to 0, NFET device is open, the PFET is closed, while the NFET for the series relationship, it is less affected by self-heating, as shown in Figure 12. Figure 12 illustrates that the temperature rise of NSFET is close to that of CFET, however, the $\Delta t_r/t_r$ of CFET is much larger, this is because the rise time of CFET is shorter than that of NSFET due to the larger saturation current, and self-heating results in the rise time variation $\Delta t_r$ of the two being close to each other, so that CFET has a larger percentage. The opposite is true for NOR, where the NFET is in parallel, resulting in a fall time that is more affected by self-heating. The principle of self-heating affecting the remaining logic gates is similar to that of NAND and NOR.

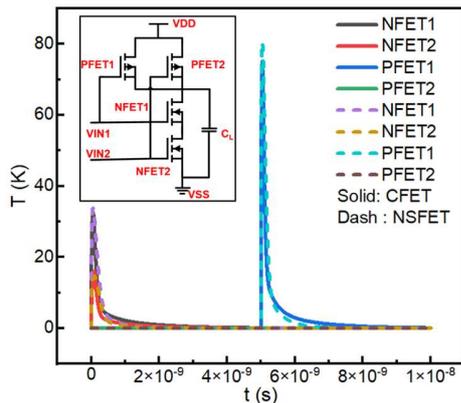

Fig.12. Time-resolved temperature rise for transistors in NAND cells consisting of NSFET and CFET. (inset) Schematic of NAND cell.

### C. Simulation Of Electrical And Thermal Characteristics Of Ring Oscillator

Finally, the effect of self-heating on the ring oscillator with different number of stages was investigated with the inverter $C_L$ set to 20fF per stage and fan-out number of 1. As shown in Fig. 13, the SHE leads to a larger period and smaller frequency of the ring oscillator. From Fig. 13(b), it can be seen that due to the small period of the CFET, the self-heating leads to a larger percentage change in the frequency of the ring oscillator consisting of CFETs. As the number of stages increases, the temperature rise of NFETs and PFETs decreases, and the effect of self-heating on the ring oscillator decreases and eventually stabilises.

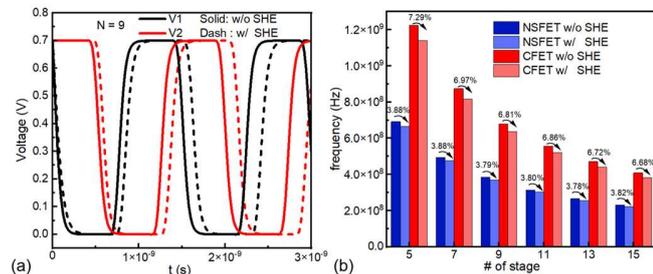

Fig.13. (a) Effect of SHE on the period of a nine-stage RO consisting of CFET, and (b) Effect of SHE on the frequency of an RO consisting of NSFET and CFET with different number of stages, with an inverter $C_L$ of 20 fF per stage.

## V. CONCLUSION

In this work, the electrical and thermal characteristics of NSFETs and CFETs are investigated, and it is found that the SHE is more significant and the on-state current degradation is larger in CFETs. Moreover, the thermal crosstalk between neighboring devices is 17% in NSFET and 32% between N/PFETs in CFET, so a thermal network model considering thermal crosstalk between neighboring devices is proposed. The electrical and thermal characteristics of inverters, logic gates and ring oscillators are further investigated in conjunction with the proposed thermal network model, and it is found that the self-heating of PFETs affects the electrical characteristics of logic gates and other electrical characteristics more than that of NFETs due to the greater on-state current degradation and higher temperature rise of PFETs. And the SHE on the transient characteristics of logic gates composed of CFETs can be up to

15.3%, and the effect on the frequency of the ring oscillator is up to 7.29%, which are both higher than that of NSFETs and should be paid extra attention to. The thermal network model proposed in this paper can be further used to study the design technology co-optimization of NSFETs and CFETs, to mitigate the self-heating of the devices and to improve the electrical performance of the devices and circuits.

## VI. ACKNOWLEDGEMENTS

This work was supported by the National Natural Science Foundation of China under Grant 62204150, the Shanghai Natural Science Foundation under Grant 23ZR1422500, and the Shanghai Science and Technology Innovation Action under Grant 22xtcx00700.